\documentstyle[12pt,epsfig]{article}
\newcommand\slv{v\kern-5pt\raise1pt\hbox{$\scriptstyle/$}\kern1pt}

\def\be{\begin{equation}}
\def\ee{\end{equation}}
\def\bq{\begin{eqnarray}}
\def\eq{\end{eqnarray}}
\newcounter{saveeqn}
\newcounter{App} 

\begin{document}
\thispagestyle{empty}
\begin{flushright}
WUE-ITP-99-017\\
NIKHEF-99-031\\
UM-TH-99-10
\end{flushright}
\vspace{0.5cm}
\begin{center}
{\Large \bf Predictions on $B \to \pi \bar{l} \nu_l $, 
$D \to \pi \bar{l} \nu_l $ 
and $D\to K \bar{l} \nu_l $ from QCD Light-Cone 
Sum Rules}\\[.3cm]
\vspace{1.7cm}
{\sc \bf A.~Khodjamirian$^{1,2,a}$, R.~R\"uckl$^{2}$, S. Weinzierl$^{3}$,  
C.W.~Winhart$^{2}$, O.~Yakovlev$^{2,4}$}\\[1cm]
$^1$ {\em The Niels Bohr Institute,  DK-2100 Copenhagen, Denmark} \\
\vspace{4mm}
$^2$ {\em Institut f\"ur Theoretische Physik, Universit\"at W\"urzburg,
D-97074 W\"urzburg, Germany} \\
\vspace{4mm}
$^3$  {\em NIKHEF, P.O. Box 41882, 1009-DB Amsterdam, The Netherlands}\\
\vspace{4mm} 
$^4$  {\em Randall Laboratory of Physics, University of Michigan,\\ Ann
Arbor, Michigan 48109-1120, USA }
\vspace{4mm}
\end{center}
\vspace{2cm}

\begin{abstract}\noindent
{The $f^+$ form factors 
of the $B\to \pi$, $D\to \pi$ and $D\to K$ transitions
are calculated from QCD light-cone sum rules (LCSR)
and used to predict the widths and differential distributions 
of  the exclusive semileptonic decays $B\to \pi \bar{l}\nu_l$,  
$D \to\pi \bar{l}\nu_l$ and $D \to K \bar{l}\nu_l$, where $l=e,\mu$. 
The current theoretical uncertainties are estimated.
The LCSR 
results are found to agree with the results of 
lattice QCD calculations and with experimental data 
on exclusive semileptonic $D$ decays.
Comparison of the LCSR prediction on $B\to \pi \bar{l} \nu_l$
with the CLEO measurement yields a value of $|V_{ub}| $ in 
agreement with other determinations.}    
\end{abstract}

\vspace*{\fill}

\noindent $^a${\small \it On leave from 
Yerevan Physics Institute, 375036 Yerevan, Armenia } \\

\newpage

\section {Introduction}

The measurement of the exclusive semileptonic 
decay $B\to \pi \bar{l} \nu_l$ by the CLEO Collaboration \cite{CLEO} 
can be used to determine
the CKM parameter $|V_{ub}|$.
This exclusive method provides an important alternative 
to the extraction of $|V_{ub}|$ from  
inclusive measurements of $B\to X_u \bar{l} \nu_l$.
However, it requires a reliable
calculation of the form factor $f^+_{B\pi}(p^2)$ 
defined by 
\be
\langle\pi(q)\mid \bar{b}\gamma_\mu u\mid B(p+q)\rangle
=2f^+_{B\pi}(p^2) q_\mu +(f^+_{B\pi}(p^2)+f^-_{B\pi}(p^2)) p_\mu ~,
\label{def}
\ee
$q$ and $p+q$ being the $\pi$- and $B$-meson 
four-momenta, respectively. 
In the case of semileptonic decays into the light leptons $l=e,\mu$, 
the form factor $f^-_{B\pi}(p^2)$ plays a negligible role.
A particularly promising approach 
to evaluate $f^+_{B\pi}(p^2)$ is based on 
QCD light-cone sum rules (LCSR) \cite{LCSR} which combine 
operator product expansion (OPE) on
the light-cone \cite{BL,ER,CZ} with QCD sum rule 
techniques \cite{SVZ}. 
The twist 2, 3 and 4 contributions to the LCSR for $f^+_{B\pi}(p^2)$ 
in leading order in $\alpha_s$ have been derived in Ref. \cite{BKR,BBKR},
while the next-to-leading order corrections to the 
twist 2 term have been calculated in 
Ref. \cite{KRWY,Bagan,KRWY2}.
The LCSR technique has further been applied to 
$B\to K$ \cite{BKR,Ball},
$D \rightarrow \pi$ \cite{BBKR,KRWY2}, and 
$D \rightarrow K$ \cite{BBD} transition form factors.

In this paper we update and improve the predictions on 
the decay distributions and integrated widths for  $B \to \pi \bar{l} \nu_l$,
$D \to \pi \bar{l} \nu_l$ and $D \to K \bar{l} \nu_l$.
In particular, we include the twist 2 next-to-leading order (NLO) 
$\alpha_s$-corrections into the calculation of the form factors 
$f^+_{D\pi}$ and $f^+_{D K}$. Moreover, we reanalyse the
momentum dependence of the form factors.
The LCSR for $f^+(p^2)$ is valid at small and intermediate momentum transfer
squared
\be
\label{chi}
p^2 \le m_Q^2 - 2 m_Q \chi ~,
\ee
where $\chi$ is a typical hadronic scale of roughly $500 \; \mbox{MeV}$
and independent of the heavy quark mass $m_Q$.
In order to go beyond this limit, we use a second LCSR for the residue 
of the pole contribution from 
the ground-state vector mesons $B^\ast$, $D^\ast$ and $D_s^\ast$, 
respectively, which
are expected to dominate at large $p^2$.
Previously, we interpolated between 
the LCSR prediction at small $p^2$ and the single-pole
approximation at large $p^2$ using a simple, but physically not very
intuitive parametrization.
In the present paper, we follow a different philosophy. 
We use the general dispersion relation for the form factor $f^+(p^2)$
and model the integral over the excited vector meson states 
by an effective pole. This yields a 
two-pole representation of $f^+$ as suggested recently in Ref. \cite{BK}.
The parameters of this representation are determined from the two light-cone
sum rules. 
This approach is more physical and has the benefit of making eventual
effects from excited vector meson states more transparent.
Finally, we discuss the sources of theoretical uncertainties of the 
LCSR method one by one, and give a careful estimate of the present overall 
uncertainty. 

Wherever possible, we compare our results with the latest lattice data.
From $B\to \pi \bar{l} \nu_l$, the CKM-matrix element $|V_{ub}|$ 
is determined by comparing the LCSR and experimental widths.
Conversely, the LCSR method is tested  
using the experimental widths of $D\to \pi \bar{l} \nu_l$ and 
$D\to K \bar{l} \nu_l$ 
and the known values of $|V_{cd}|$ and $|V_{cs}|$, respectively. 
This analysis favors a value $m_s$(1 GeV) 
$\approx$ 150 MeV for  the $s$-quark mass.

The paper is organized as follows. 
Sect. 2 and 3 are devoted to the $B \rightarrow \pi$
transition as a prototype example.
In Sect. 2 we present the LCSR analysis of $f^+_{B\pi}(p^2)$,
the theoretical uncertainties
of which are estimated and discussed in Sect. 3.
The analogous analysis of the $D \rightarrow \pi$ and $D \rightarrow K$
transitions is described in Sect. 4.
Sect. 5 deals with applications to decay distributions and  
integrated widths for  $B \to \pi \bar{l} \nu_l$, 
$D\to \pi \bar l \nu_l$ and $D\to K \bar l \nu_l$. This section
also summarizes the comparison of theory with experiment.

\section{The form factor $f^+_{B\pi}(p^2)$}  

The LCSR for the form factor $f^+_{B\pi}(p^2)$ is obtained from the
correlation function
\bq
\label{corrfct}
F_\mu(p,q) & = & i \int dx e^{i p \cdot x} \langle \pi(q) |
T \left\{ \bar{u}(x) \gamma_\mu b(x), m_b \bar{b}(0) i\gamma_5 d(0)
\right\} | 0 \rangle
\eq 
by contracting the $b$-quark fields in the time-ordered product of currents,
expanding the remaining matrix elements of nonlocal operators in terms
of light-cone distribution amplitudes of the pion, and 
writing a dispersion relation in the $B(\bar{b}d)$-channel.
The derivation is described in detail in Ref. \cite{BKR,BBKR,KRWY,KR}.
Schematically, the resulting sum rule has the form
\bq\label{fplus}
f^+_{B\pi}( p^2)&=&\frac{1}{2m_B^2f_B}
\exp \left( \frac{m_B^2}{M^2}\right)
\Bigg [ F_0^{(2)}(p^2,M^2,m_b^2,s_0^B,\mu_b) \\ \nonumber
&+&\frac{\alpha_s(\mu_b)}{3\pi} F_1^{(2)}(p^2,M^2,m_b^2,s_0^B,\mu_b)
+ F_0^{(3,4)}(p^2,M^2,m_b^2,s_0^B,\mu_b) \Bigg ]~,
\eq
where $m_B$ is the $B$-meson mass, $m_b$ the $b$-quark pole mass, 
and $f_B$  the $B$-meson decay constant defined by the matrix element
\bq
\langle 0 \mid m_b\bar{q}i\gamma_5 b\mid B\rangle =m_B^2f_B~.
\label{fB2}
\eq
The mass scale $M$ is associated with a Borel transformation 
usually performed in sum rule calculations.  It characterizes 
the off-shellness of the $b$-quark. The scale $\mu_b$ 
is the factorization scale separating soft and hard dynamics.
Long-distance effects involving scales lower than
$\mu_b$ are absorbed in the pion distribution amplitudes which represent the
universal nonperturbative input in LCSR. They have been studied up to twist 
4 and are given in Ref. \cite{BBKR,KR,BF}.
The short-distance effects are incorporated in hard-scattering amplitudes 
calculated perturbatively and convoluted with the pion distribution amplitudes.
The first two terms of the bracket in (\ref{fplus}) represent the
NLO twist 2 contributions, while the third term refers to the
twist 3 and 4 contributions which are only known in LO.
For illustration, the leading term $F_0^{(2)}$ is given by
\bq
\label{twist2}
F_0^{(2)}(p^2,M^2,m_b^2,s_0^B,\mu_b) & = &
m_b^2 f_{\pi} \int\limits_{\Delta}^1\frac{du}{u} 
exp\left(-\frac{m_b^2-p^2(1-u)}{uM^2} \right)\varphi_\pi(u,\mu_b)
\eq
with $f_\pi = 132 \; \mbox{MeV}$ being the decay constant
and $\varphi_\pi$ being the twist 2 distribution amplitude
of the pion. 
The latter can be interpreted as the probability
amplitude for finding a quark with momentum fraction $u$
inside a pion.
The lower integration boundary $\Delta=(m_b^2-p^2)/(s_0^B-p^2)$
is determined by the
effective threshold parameter $s_0^B$ which originates from
the subtraction of excited resonances and continuum states 
contributing to the dispersion integral in the $B$ 
channel. This subtraction is performed assuming quark-hadron duality
at $(p+q)^2 \ge s_0^B$.
The explicit expressions for the remaining terms $F_1^{(2)}$ and
$F_0^{(3,4)}$ can be found in Ref. \cite{KRWY,Bagan} and
Ref. \cite{BKR,BBKR}, respectively.

\begin{figure}
\centerline{
\epsfig{file=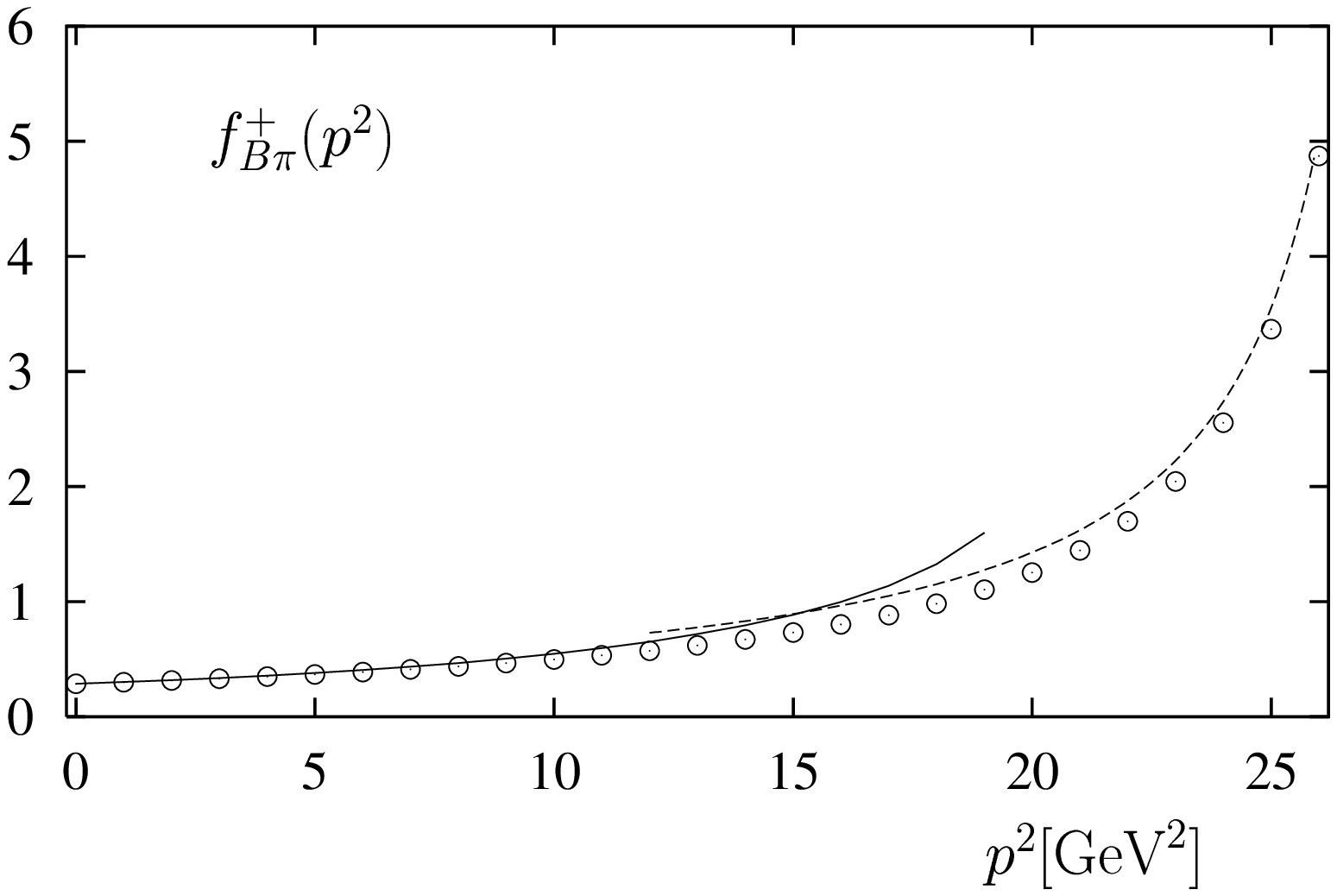,scale=0.8,%
clip=}}
\caption{\label{fig1} \it The $B\to\pi$ form factor: direct 
LCSR prediction (solid curve),  
$B^*$-pole contribution with the LCSR estimate of the residue
(dashed curve), and two-pole parametrization (\ref{paramB}) (circles).}
\end{figure}

Numerically, we take $f_B=180 \pm 30$ MeV, $m_b=4.7 \mp 0.1$ GeV, 
$s_0^B=35 \pm 2$ GeV$^2$, and  
$\mu_b= \sqrt{m_B^2-m_b^2}\approx 2.4 $ GeV.
Here and in forthcoming theoretical results, the error notation 
is to be interpreted as a
range reflecting the present theoretical uncertainty, e.g.,
$f_B=150 \div 210$ MeV.
It is also important to note that
the above parameters are interrelated by the two-point QCD sum rule for 
$f_B$ \cite{fb}. Consequently, their variation within the given
ranges is correlated 
as indicated by the alternating $\pm$ signs. Furthermore,  
sum rules for observables should in principle be independent of 
the auxiliary Borel parameter $M^2$. In practice, however, this is 
not the case because of the various approximations made. 
The allowed range of $M^2$ differs for different 
sum rules. For LCSR  it is usually determined by requiring 
the twist 4 contribution not to exceed $10\%$ and 
the contributions from excited and continuum states 
to stay below $30\%$. Specifically, for the sum rule (\ref{fplus}),
these criteria yield $M^2 = 10 \pm 2 \; \mbox{GeV}^2$.
In the case of the two-point sum rule for $f_B$,
the allowed interval of the Borel parameter is 
$M^2 = 4 \pm 2 \; \mbox{GeV}^2$.
With the nominal
 values of the parameters specified above   
the LCSR (\ref{fplus}) leads to 
the form factor $f^+_{B\pi}(p^2)$ shown by the solid curve in Fig. \ref{fig1}.
In particular, at $p^2=0$ one gets 
\begin{eqnarray}\label{nfplus}
\quad~ f^+_{B\pi}(0) = 0.28 \pm 0.05~.
\end{eqnarray}
The estimate of the theoretical uncertainty will be explained in detail 
in the following section.

As already mentioned in the introduction,
the LCSR (\ref{fplus}) is expected to hold only at
$p^2 \le m_b^2 -2 m_b \chi \approx 18 ~\mbox{GeV}^2$.
Indeed, at $p^2 > 20 \; \mbox{GeV}^2$ the twist 4 
contribution is found to grow strongly, and the 
stability of the sum rule against variation of the Borel parameter
$M^2$ is lost. This clearly signals the breakdown of the light-cone expansion.
On the other hand, as $p^2$ approaches the kinematical limit
$(m_B - m_\pi)^2$ the lowest-lying $B^\ast$ pole is expected to give 
the dominant contribution to $f^+_{B\pi}$.  
The residue of this pole contribution is given by the product of 
the $B^*$ decay constant defined by
\be\label{disG}
\langle 0 \mid \bar{q}\gamma_\mu b\mid B^*\rangle
=m_{B^*}f_{B^*}\epsilon_\mu ~,
\label{fB*}
\ee
and the strong $B^*B\pi$ coupling constant defined by
\begin{equation}
\langle \bar{B}^{*0}\pi^-\mid B^-\rangle =
-g_{B^*B\pi}(q \cdot\epsilon )~.
\label{BstarBpi}
\end{equation}
This product can be calculated from another LCSR which
follows from the same
correlation function (\ref{corrfct}) as the LCSR (\ref{fplus}), considering
this time, however, a 
double dispersion relation in the $B$ and $B^\ast$ channel.
Again, we only indicate the schematic form of this 
sum rule \cite{BBKR,KRWY2}:
\begin{eqnarray}
f_{B^*}g_{B^*B\pi}&=&\frac{1}{m_B^2m_{B^*} f_B}
e^{\frac{m_{B}^2+m_{B^*}^2}{2M^2}}
\Bigg[ G_0^{(2)}(M^2,m_b^2,s_0^B,\mu_b)
\nonumber
\\
& & +\frac{\alpha_s(\mu_b)}{3 \pi}G^{(2)}_1(M^2,m_b^2,s_0^B,\mu_b)
+ G_0^{(3,4)}(M^2,m_b^2,s_0^B,\mu_b)\Bigg]~. 
\label{srcoupl}
\end{eqnarray}
In analogy to (\ref{fplus}), the NLO twist 2 contributions are denoted by
$G_0^{(2)}$ and $G_1^{(2)}$, while $G_0^{(3,4)}$ stands for the LO twist
3 and 4 contribution. Explicitly, the leading twist 2 term is given by 
\bq
G_0^{(2)}(M^2,m_b^2,s_0^B,\mu_b)
& = &
m_b^2 M^2 \left(e^{-\frac{m_b^2}{M^2}} - 
e^{-\frac{s_0^B}{M^2}}\right)f_\pi\varphi_\pi(u_0,\mu_b)~.
\label{twist2coupl}
\eq
In contrast to (\ref{twist2}) where one has
an integral over the normalized distribution amplitude $\varphi_\pi$,
the above term depends on the value of 
$\varphi_\pi$ at the point $u_0 \approx 0.5$. This difference also applies
to terms of higher twist, whence the LCSR (\ref{srcoupl}) is much more 
sensitive to the precise shape of the pion distribution amplitudes 
than the LCSR (\ref{fplus}).
The parameters of the two LCSR coincide with the 
exception of the Borel mass which in the case of (\ref{srcoupl}) is constrained
to the interval $M^2 = 9 \pm 3 \; \mbox{GeV}^2$.
Numerically, we obtain \cite{KRWY2} 
\be\label{fg}
f_{B^{*}}g_{B^*B\pi }=  4.4 \pm 1.3  \,\mbox{\rm GeV} ~.
\label{combinB}
\ee
Again, the uncertainty estimate will be discussed in the next section.

In order to determine the form factor $f^+_{B\pi}(p^2)$ 
at large momentum transfers $p^2 > 18 $ GeV$^2$, where we cannot rely 
on the LCSR (\ref{fplus}), we consider the dispersion relation 
\be
f^+_{B\pi}(p^2) = \frac{f_{B^*}g_{B^*B\pi}}{2m_{B^*}(1-p^2/m^{2}_{B^*})}+
\int\limits_{s_0}^{\infty}\frac{d\tau \rho(\tau)}{\tau-p^2}~.
\label{vectpole}
\ee
Here, the pole term is due to the ground state $B^*$ meson
and the dispersive integral takes into account contributions 
from higher resonances and continuum states in the $B^\ast$ channel.
Using (\ref{combinB}), 
the $B^*$-pole contribution is shown by the dashed
curve in Fig. \ref{fig1}.
With decreasing momentum transfer the one-pole approximation
deviates noticeably from the LCSR result (\ref{fplus}).
At $p^2 = 0$ the difference reaches about $50\%$ 
showing that the dispersion integral in (\ref{vectpole})
over the heavier states cannot be neglected.
In Ref. \cite{BK}, it has been suggested to model their contribution 
by an effective second pole: 
\be\label{pBK}
f_{B\pi}^+(p^2)=c_B\Big(\frac{1}{1-p^2/m^2_{B^*}}
-\frac{\alpha_{B\pi}}{1-p^2/\gamma_{B\pi}m^2_{B^*}}\Big) ~. 
\ee
By means of the LCSR (\ref{fplus}) and (\ref{srcoupl}) one can now determine
the parameters $c_B$ and $\alpha_{B\pi}$.
From (\ref{fg}) and (\ref{vectpole}) we get
\be
c_B= \frac{f_{B^*}g_{B^*B\pi}}{2m_{B^*}}=0.41 \pm 0.12~, 
\label{cb}
\ee
and putting $p^2=0$ in (\ref{pBK}) and using directly 
(\ref{fplus}) and (\ref{srcoupl})
we obtain 
\be\label{alpha}
\alpha_{B\pi} 
= 1- \frac{2m_{B^*} f_{B\pi}^+(0)}{f_{B^*}g_{B^*B\pi}} 
=  0.32 ^{+0.21}_{-0.07}~. 
\ee
It should be emphasized that the latter result is independent of $f_B$
or the corresponding two-point sum rule. Moreover, since the LCSR 
(\ref{fplus}) and (\ref{srcoupl}) involve common parameters, 
some of the uncertainties cancel in the ratio (\ref{alpha}).
In the heavy quark limit \cite{KR},
$f_{B\pi}^+(0)$ should scale like $1/m_b^{3/2}$ which    
implies a positive sign for $\alpha_{B\pi}$.
Thus, the result (\ref{alpha}) nicely
demonstrates the consistency of the LCSR method with the heavy quark limit.
The remaining parameter $\gamma_{B\pi}$ can in principle be obtained by 
fitting (\ref{pBK}) to (\ref{fplus}) at $p^2 < 15~ \mbox{GeV}^2$ 
with $c_B$ and $\alpha_{B\pi}$ fixed. However, since the 
LCSR prediction deviates in shape
very little from the $B^\ast$-pole contribution as can be anticipated 
from Fig. \ref{fig1}, the fit only gives the lower bound $\gamma_{B\pi} > 2$.

In the combined limit $m_Q \to \infty$ and $E_{\pi} \to \infty$, 
one has a relation between $f_{B\pi}^+$ and the scalar form factor 
\bq
f^0_{B\pi}(p^2) & = & f^+_{B\pi}(p^2)
+ \frac{p^2}{m_B^2-m_\pi^2} f^-_{B\pi}(p^2) ~,
\eq
namely \cite{Charles}
\bq
f^0_{B\pi} & = & \frac{2 E_{\pi}}{m_B} f^+_{B\pi}.
\eq
If a parametrization similar to the second term of (\ref{pBK})
is used for $f^0_{B\pi}$, this suggests $\gamma_{B\pi} =1/\alpha_{B\pi}$
\cite{BK}.
Interestingly, the fit described above is well consistent 
with this constraint. Therefore, we will assume this relation in the following.

Our final result for the   
form factor $f^+_{B\pi}(p^2)$ can then be written in a very convenient
form:
\be\label{paramB}
f_{B\pi}^+(p^2)=\frac{f_{B\pi}^+(0)}{(1-p^2/m_{B^*}^2)(1-
\alpha_{B\pi}p^2/m_{B^*}^2)} ~, 
\ee
where $f^+_{B\pi}(0) = c_B ( 1 - \alpha_{B\pi})$
has been used, and 
$f^+_{B\pi}(0)$ and $\alpha_{B\pi}$ are given in 
(\ref{nfplus}) and (\ref{alpha}), respectively.
The above parametrization is plotted in Fig. \ref{fig1}. 
We see that it coincides nicely with the LCSR prediction (\ref{fplus})
at low $p^2$ and approaches the single-pole approximation at large $p^2$.
Fig. 1 therefore shows that at small and intermediate momentum transfer our 
result is actually model-independent the two-pole model (\ref{pBK})
being nothing but a convenient parametrization in this region.

\begin{figure}
\centerline{
\epsfig{file=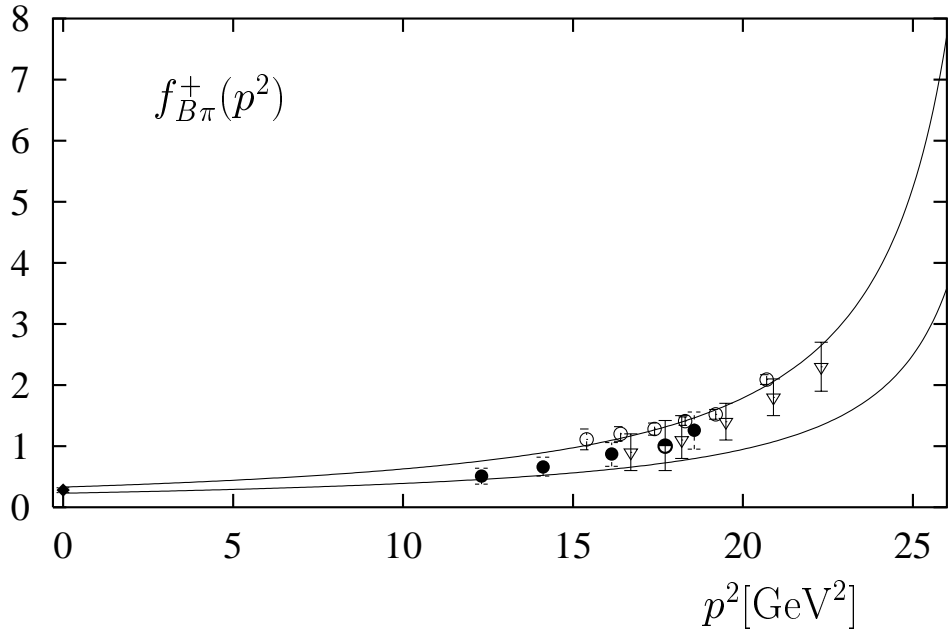,scale=0.8,%
clip=}}
\caption{\label{fig2} \it The LCSR prediction for the $B\to\pi$  
form factor in comparison to lattice results. 
The full curves indicate the size of the LCSR uncertainties.
The lattice results come from FNAL \cite{Flynn} (full circles),
UKQCD \cite{UKQCD} (triangles), APE \cite{APE} (full square), 
JLQCD \cite{JLQCD} (open circles), and ELC \cite{Flynn} (semi-full circle).}
\end{figure}

\begin{figure}
\centerline{
\epsfig{file=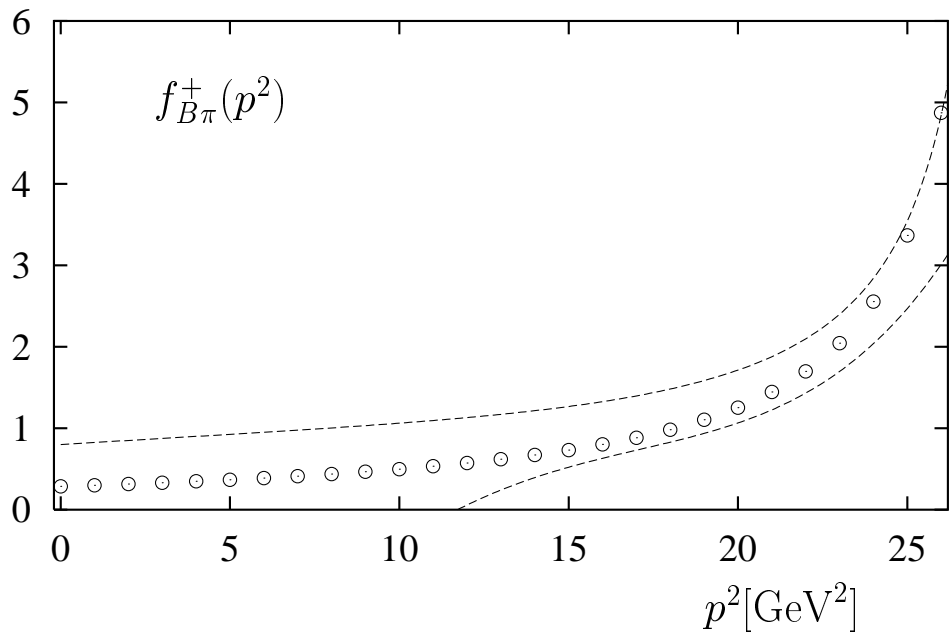,scale=0.8,%
clip=}}
\caption{\label{fig3} \it 
The LCSR prediction on the form factor $f_{B\pi}^+$ (circles)
in comparison to the constraint (dashed) derived in 
Ref. \cite{Boyd}.} 
\end{figure}

Fig. \ref{fig2} shows a comparison of (\ref{paramB}) with
recent lattice results \cite{Flynn,lattrev, UKQCD,APE,JLQCD}.
The agreement within uncertainties is very satisfactory.
Here, the overall uncertainty in the LCSR prediction is estimated by 
adding the uncertainties from individual sources linearly. 
This explains why the range of uncertainty updated in Fig. \ref{fig2}
is larger than the uncertainty estimated previously by us and other authors
adding the various uncertainties in quadrature.
We consider the present procedure to be the appropriate treatment
of theoretical uncertainties. Finally, 
the LCSR prediction also obeys the constraints
derived from sum rules for the inclusive
semileptonic decay width in the heavy quark limit \cite{Boyd}.
This is demonstrated in Fig. \ref{fig3}.

\section{Theoretical uncertainties}

The theoretical uncertainties in the LCSR (\ref{fplus}) and
(\ref{srcoupl}) originate from uncertainties in the input
parameters and from unknown contributions of higher order 
in twist and $\alpha_s$.
The former are estimated by varying the numerical values of the input
parameters within the ranges given in Sect. 2, 
for the latter we present some plausible arguments concerning the 
size of these corrections.
If not 
stated otherwise only a single parameter is varied at a time, 
while the other parameters are held fixed. 
For $f_B$ we substitute the corresponding two-point sum rule.
The uncertainty in a given quantity is 
then expressed by plus/minus the interval of variation w.r.t.
the result obtained for the nominal values of the input parameters.
Although in the parametrization (\ref{paramB})  
of $f^+_{B\pi}(p^2)$ the LCSR (\ref{fplus}) is only used to 
determine the normalization 
at $p^2=0$, we investigate the uncertainty of
the LCSR prediction (\ref{fplus})
in the whole range of validity, that is at $0 \le p^2 < 18 ~\mbox{GeV}^2$.
This serves as a cross-check and ensures the consistency
of (\ref{paramB}) with the LCSR (\ref{fplus}) in the whole range of overlap.
In the region $p^2 > 18 \;\mbox{GeV}^2$ we rely on the parametrization
(19) in a more substantial way. This may introduce some 
model-dependence due to the particular functional form assumed. 
However, since only the 
kinematically suppressed region of large momentum transfer 
is affected, this uncertainty 
has little influence on the integrated width and the value 
of $V_{ub}$ extracted from the latter.
Our findings are summarized below:

(a) Borel mass parameter

\begin{figure}
\hspace {-2cm}
\centerline{
\epsfig{file=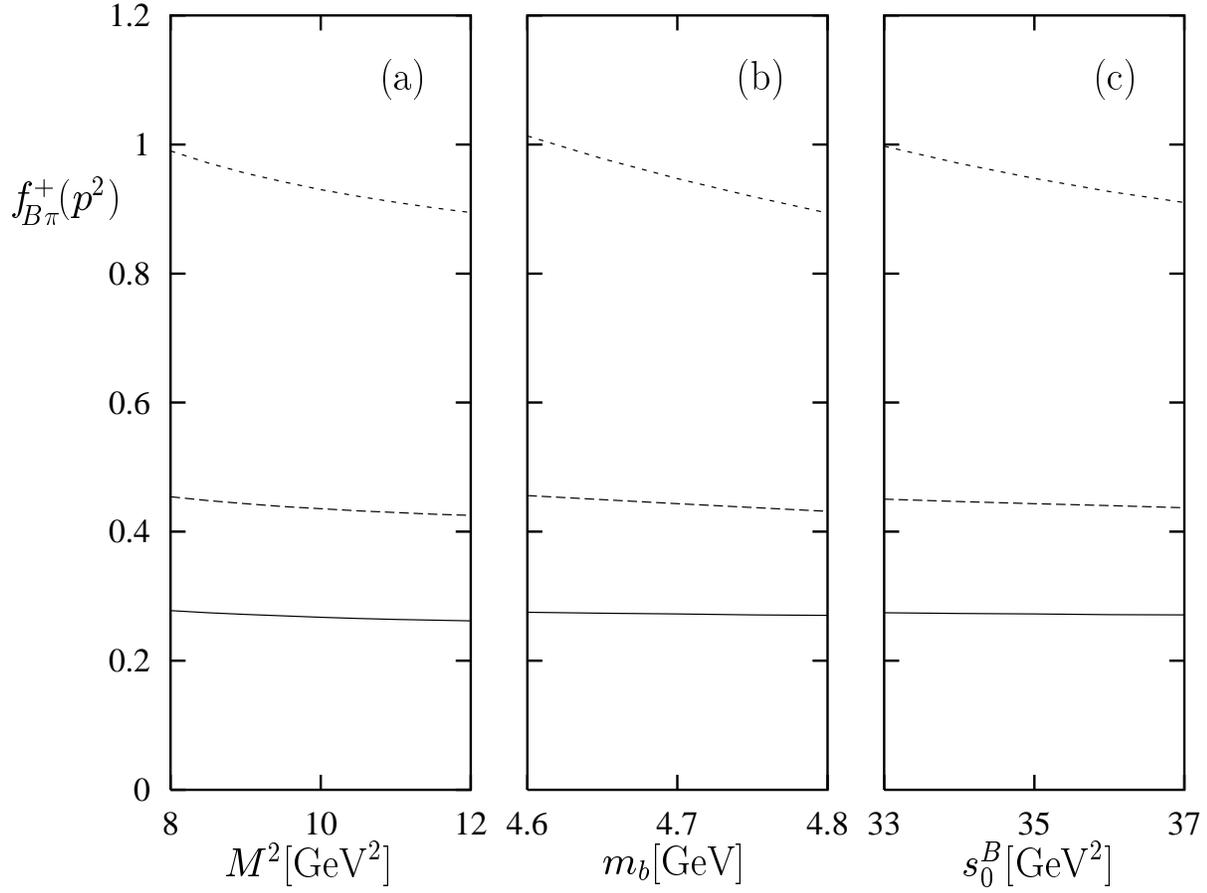,scale=0.9,%
clip=}}
\vspace {-12cm} 
\caption{\label{fig4} \it Sensitivity
of the LCSR for $f_{B\pi}^+(p^2)$ 
on the Borel parameter (a), the $b$-quark mass (b),
and the subtraction threshold (c)
at momentum transfer squared $p^2=0$ (solid),
$p^2=8$ GeV$^2$ (long-dashed), and $p^2=16$ GeV$^2$ (short-dashed).}
\end{figure}

\noindent The variation of $f^+_{B\pi}$ with $M^2$ is illustrated in 
Fig. \ref{fig4}a. 
It turns out to be rather small, $\pm (3 \div 5)\%$ depending on $p^2$.
The corresponding variation of $f_{B^*}g_{B^*B\pi}$ amounts to $\pm 10\%$. 

(b) $b$-quark mass and subtraction threshold  

\noindent Fig. \ref{fig4}b and \ref{fig4}c 
show the variation of $f^+_{B\pi}$ with $m_b$
and $s_0^B$, respectively.  
If $m_b$ and $s_0^B$ are varied simultaneously such that 
one achieves maximum stability of the sum rule for $f_B$, 
the change in $f^+_{B\pi}$ is negligible at 
small $p^2$ rising to about $\pm 3\%$ at large $p^2$.
The corresponding variation of $f_{B^*}g_{B^*B\pi}$ is about $\pm 4\%$.

(c) quark condensate density

\noindent The coefficient $\mu_\pi=m_\pi^2/(m_u+m_d)$ of the 
twist 3 pion distribution amplitude is related to the quark condensate
density $\langle \bar q q \rangle$ by PCAC. 
Therefore, the uncertainty in the quark condensate  
$\langle \bar q q \rangle(\mu_b)= -(268\pm 10 ~\mbox{MeV})^3$  
induces an uncertainty in both the sum rule for $f_B$ 
and the terms $F^{3}_0$ and  $G^{3}_0$ of the 
LCSR (\ref{fplus}) and (\ref{srcoupl}), respectively. The resulting  
uncertainty on $f^+_{B\pi}$ and 
$f_{B^*}g_{B^*B\pi}$ is about $\pm3\%$.
Gluon and quark-gluon condensates have little influence on $f_B$ 
and no direct connection to the LCSR considered here. 

(d) higher-twist contributions

\begin{figure}[p]
\centerline{
\epsfig{file=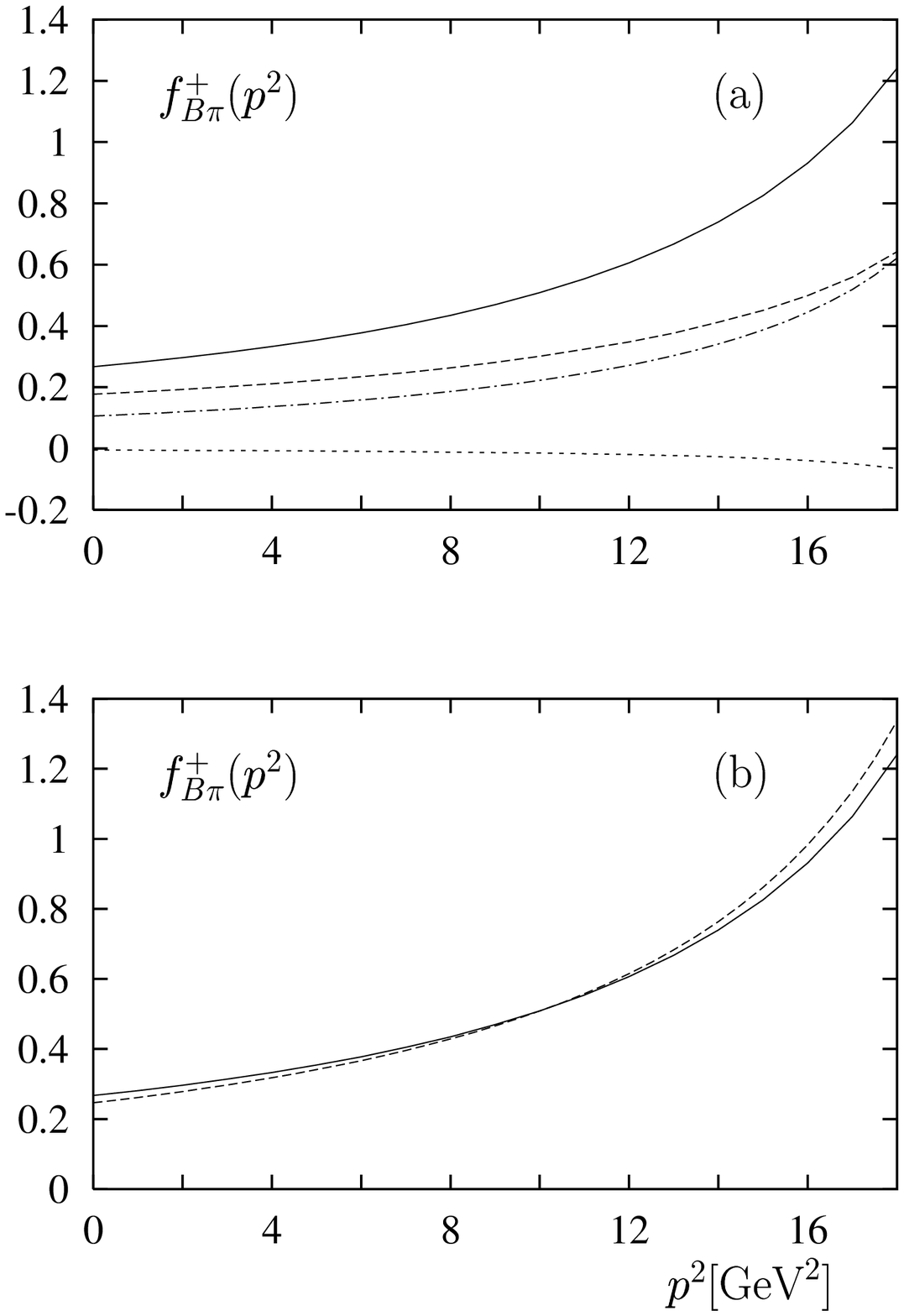,scale =0.6,width=\textwidth,bbllx=0pt,bblly=80pt,
bburx=600pt,%
bbury=800pt,clip=}
} 
\caption{\label{fig5} \it  The LCSR prediction on $f_{B\pi}^+$: 
(a) individual contributions
from twist 2 (dashed), 3 (dash-dotted), 4 (dotted), 
and the total sum (solid); 
(b) with (solid) and without (dashed) 
non-asymptotic corrections to the pion distribution amplitudes.}
\end{figure}

\noindent No reliable estimates exist for distribution amplitudes 
beyond twist 4. Therefore, we use the magnitude of the twist 4 
contribution to $f^+_{B\pi}$
as an indicator for the uncertainty due to the neglect of 
higher-twist terms. From Fig. \ref{fig5}a we see that 
the twist 4 term of (\ref{fplus}) contributes
less than 2~\% at low
$p^2$ and about 5~\% at large $p^2$ to $f^+_{B\pi}$. 
Also the twist 4 term in (\ref{srcoupl}) contributes no more than 5\%
to $f_{B^*}g_{B^*B\pi}$.

(e) pion distribution amplitudes 

\noindent
The asymptotic distribution amplitudes and the scale dependence 
of the nonasymptotic coefficients are determined by 
perturbative QCD. 
However, the values of the nonasymptotic 
coefficients at a certain scale $\mu_0$ are 
of genuinely nonperturbative origin.
They can be determined either from experiment or, eventually, 
from lattice QCD 
\cite{nonasympt}. 
For illustration, the twist 2 distribution amplitude
appearing in (\ref{twist2}) and (\ref{twist2coupl}) is given by
\begin{equation}
\varphi_\pi(u,\mu) = 6 u(1-u)\Big[1+a_2^{\pi}(\mu)C^{3/2}_2(2u-1)+
 a_4^{\pi}(\mu)C^{3/2}_4(2u-1)]~,
\label{expansion}
\end{equation}
where $C_{n}^{3/2}(x)$ are Gegenbauer polynomials, 
and $a_n^{\pi}(\mu)$ are the nonasymptotic coefficients.
Investigation by means of conformal partial wave expansion
justifies the neglect of terms with $n > 4$
(see, e.g., Ref. \cite{BallDA} for further explanation and references). 

In Ref. \cite{BBKR,KRWY,KRWY2} we have   
used the Braun-Filyanov (BF) distribution amplitudes \cite{BF}.
Two recent analyses based on the LCSR for the 
$ \gamma^* \gamma \to \pi^0 $ transition form factor \cite{AK,SY}
and the pion form factor \cite{BKM} indicate 
that nonasymptotic effects in $\varphi_\pi$ 
are in fact smaller than the effects implied by the original
BF coefficients \cite{BF}. 
However, the uncertainties are still sizeable.
The latter is even more so for the    
nonasymptotic coefficients of the twist 3 and 4 
distribution amplitudes \cite{BallDA}. 
There is a crude, but simple and as we will see sufficient way to 
estimate the sensitivity of the LCSR to nonasymptotic effects, that
is by comparing the results obtained with BF and purely asymptotic
distribution amplitudes.
For $f^+_{B\pi}(p^2)$, this comparison is displayed in Fig. \ref{fig5}b. 
We see that the difference is very moderate:  about -7\% 
at small $p^2$ and +7\% at large $p^2$.
The intermediate region around $p^2 = 10$ GeV$^2$ is almost unaffected.
A similar investigation of (\ref{srcoupl}) shows that 
$f_{B^*}g_{B^*B\pi}$ increases by about 8 \% if all nonasymptotic
effects are disregarded.
Since the LCSR (\ref{fplus}) involves
convolutions of relatively smooth coefficient functions with 
normalized distribution amplitudes,  
the moderate sensitivity to the precise shape of the
latter is easy to understand. In contrast, the LCSR (\ref{srcoupl})
depends on the amplitudes at a given point and could, therefore,
be strongly affected by nonasymptotic effects. However, in this case
the effects have opposite signs for twist 2 and 3, and thus tend to cancel. 
 
(f) perturbative corrections 

\noindent The NLO QCD corrections
to the twist 2 contribution to $f_Bf^+_{B\pi}$ 
derived from (\ref{fplus})  
\cite{KRWY,Bagan} amount to about $(20\div30)\%$.
Corrections of similar size affect
the two-point sum rule for $f_B$ \cite{fb}.
Hence, in the ratio giving $f^+_{B\pi}$ they almost cancel leaving
a net correction of less than 10\%.
A similar cancellation takes place between the NLO corrections to
$f_Bf_{B^*}g_{B^*B\pi}$ derived from (\ref{srcoupl}) and the 
NLO corrections to
$f_B$. Here, the net effect is only 5\%.

The perturbative corrections to the higher-twist
terms are still unknown. 
Most important are the NLO corrections to $F_0^{(3)}$ and $G_0^{(3)}$.  
Optimistically, they could be as small as  
the correction to the quark condensate term
in the sum rule for $f_B$, namely about 2\% \cite{Neubert}. 
More conservatively, they may be of the same order as the 
twist 2 NLO corrections. Since in LO the twist 3 terms contribute 
about $(30\div50)\%$ to the LCSR, one may expect corrections to    
$f^+_{B\pi}$ and $f_{B^*}g_{B^*B\pi}$ as large as $(5\div15)\%$ in total.
Therefore, it is very important to make every effort to
calculate these corrections.

(g) normalization scale 

\begin{figure}
\centerline{
\epsfig{file=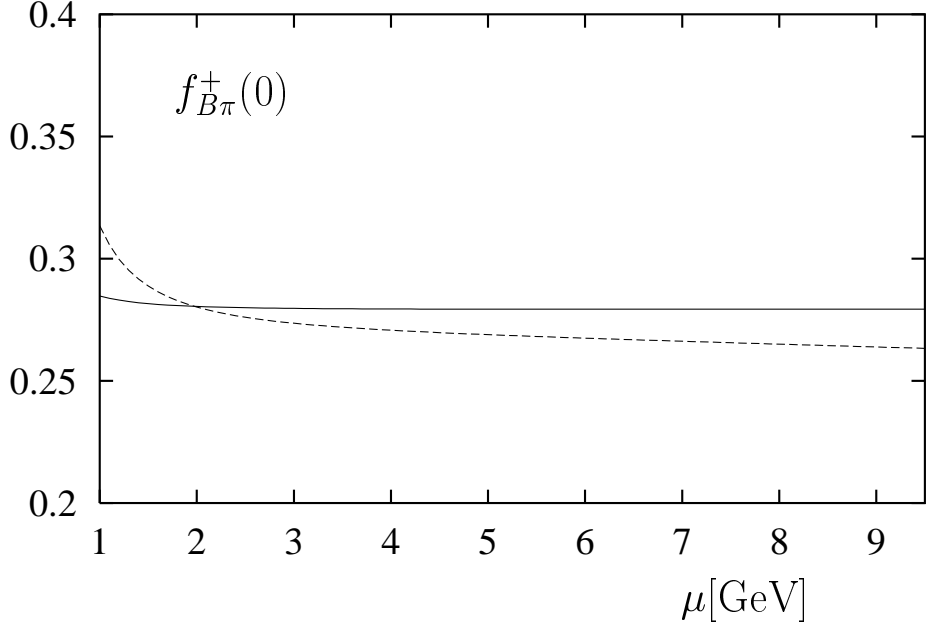,scale=0.8,%
clip=}}
\caption{\label{fig6} \it Scale dependence of
the form factor $f_{B\pi}^+(0)$ from LSCR with
$f_B$ calculated from the corresponding  $\mu$-dependent two-point sum rule
(solid), and with fixed $f_B = 180\; \mbox{MeV}$ (dashed).}
\end{figure}

\noindent Also the $\mu$-dependence of the sum rules for 
$f_B$ and $f_Bf^+_{B\pi}$ turns out to be quite similar. 
As a result, the ratio of these sum rules yielding
$f^+_{B\pi}$ shows very little scale dependence as can be seen
from Fig. \ref{fig6}. An analogous cancellation of scale dependences
is observed in the ratio of sum rules giving $f_{B^*}g_{B^*B\pi}$.

The total theoretical uncertainty 
in (\ref{fplus}) and (\ref{srcoupl}) is obtained by adding the uncertainties
from the different sources (a)-(g) linearly. However, expecting that the
twist 3 NLO corrections will be calculated in near future and assuming 
that they will turn out to be on the lower side of the range considered in
(f) we have not included them in the numerical uncertainty estimates quoted in 
this paper. The same procedure has been followed in estimating
the uncertainty on $\alpha_{B\pi}$ which is given by the ratio
of the LCSR (\ref{fplus}) and (\ref{srcoupl}), and has therefore to
be studied separately. Finally, in the parametrization
(\ref{paramB}) of the form factor $f^+_{B\pi}$ the uncertainty
in normalization and shape is given by the uncertainty in $f^+_{B\pi}(0)$
and $\alpha_{B\pi}$, respectively.

The theoretical uncertainties quoted in the following section
dealing with $D\to \pi$ and $D\to K$ transitions are obtained
analogously to the uncertainty on $B\to \pi$.

\section{ The $D\to \pi$ and $D\to K$ form factors}

With the LCSR (\ref{fplus}) and (\ref{srcoupl})
at hand it is straightforward 
to obtain the corresponding sum rules for the $D\to\pi$ 
form factor $f^+_{D\pi}$ and for the residue 
$f_{D^*}g_{D^*D\pi}/2m_{D^*}$ of the $D^*$-pole contribution.  
The input parameters are as follows. For the $c$-quark pole mass $m_c$, 
the subtraction threshold $s_0^D$, and the factorization scale 
$\mu_c$ we take  
$m_c=1.3\mp 0.1$ GeV, $s_0^D=6 \pm 1 $ GeV$^2$, and 
$\mu_c= \sqrt{m_D^2-m_c^2}\approx 1.3 $ GeV. 
The decay constant $f_D$ is calculated  
from the two-point QCD sum rule in NLO \cite{fb} with the Borel mass squared
$M^2 = 1.5 \pm 0.5$ GeV$^2$ yielding $f_D= 200 \pm 20$ MeV. 
Again, the variation of the 
parameters in the above ranges is correlated as indicated 
by the alternating $\pm$ signs.  
Furthermore, the pion distribution amplitudes are to be taken
at the scale $\mu_c$. Details can be found in Ref. \cite{KR}. 

\begin{figure}
\centerline{
\epsfig{file=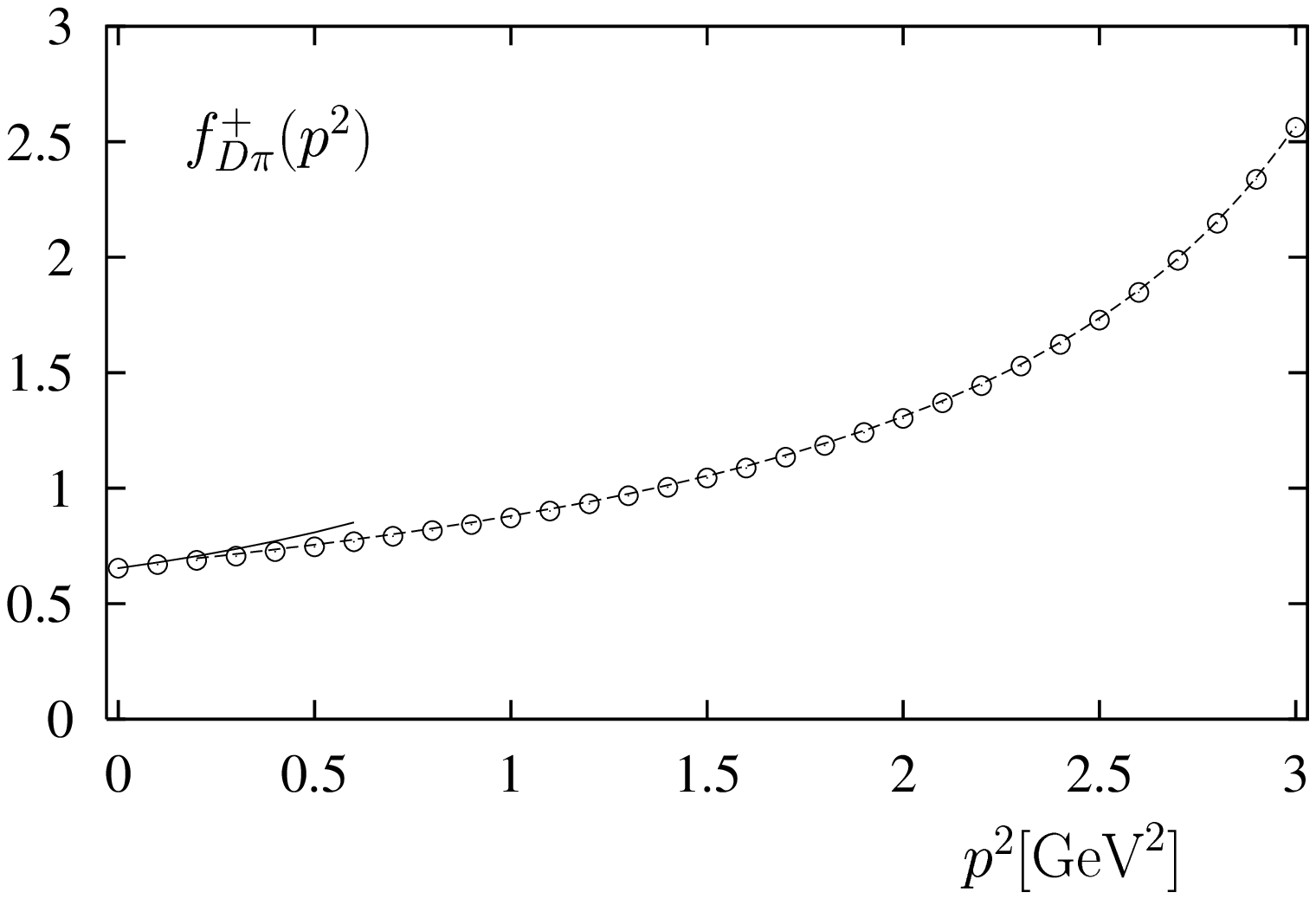,scale=0.8,%
clip=}}
\caption{\label{fig7} \it  The $D\to\pi$ form factor: direct 
LCSR prediction (solid curve),  
$D^*$-pole contribution with the LCSR estimate of the residue
(dashed curve), and two-pole parametrization (\ref{paramD}) (circles).}
\end{figure}

The form factor $f^+_{D\pi}$ resulting from (\ref{fplus}) with the
input specified above is 
shown by the solid curve in Fig. \ref{fig7}. For this illustration, we have
chosen the nominal values of the parameters. 
The range of the Borel parameter $M^2$ in which (\ref{fplus})
fulfills the criteria
on the size of the twist 4 terms and the contribution from 
excited states, is found to be $M^2 = 4 \pm 1$ GeV$^2$.
For later use, we also quote the value of $f^+_{D\pi}$ at 
zero momentum transfer:
\be
f^+_{D\pi}(0) = 0.65 \pm 0.11 
\label{fD0}
\ee 
which nicely agrees 
with lattice estimates, for example, the world average \cite{lattrev} 
\be
f^+_{D\pi}(0)= 0.65 \pm 0.10 ~,
\label{latfD0}
\ee
or the most recent APE result \cite{APE}, 
$f^+_{D\pi}(0)= 0.64 \pm 0.05 ^{+.00}_{-.07}$.

In the case of $D \rightarrow \pi$ transitions, the LCSR (\ref{fplus}) 
is reliable as long as $p^2 < m_c^2 -2m_c\chi \approx 0.6 $ GeV$^2$. 
Here, the same hadronic scale 
$\chi \approx 500$ MeV is assumed as for $B \rightarrow \pi$.
At larger momentum transfer, one may make use 
of a dispersion relation in analogy to (\ref{vectpole}). 
The residue of the pole contribution from the ground state $D^*$ meson 
can be calculated from
the LCSR (\ref{srcoupl}) adjusted to the $D^*D\pi$ coupling.
One gets
\be
f_{D^{*}}g_{D^*D\pi }= 2.7 \pm 0.8 ~\mbox{\rm GeV} 
\label{combinD}
\ee
with the allowed interval of the Borel parameter being  
$M^2 = 3 \pm 1$ GeV$^2$.
In Fig. \ref{fig7}, the $D^*$-pole contribution is
shown by the dashed curve, again for the nominal values of the parameters.
Comparison with the LCSR prediction at low $p^2$ indicates that there
is little room for additional contributions from higher resonances,
in contrast to what we have found for the $B \to \pi$ transition.

In order to quantify this statement, it is useful to investigate a
parametrization of $f^+_{D\pi}(p^2)$ similar to (\ref{paramB}):
\begin{equation}\label{paramD}
f_{D\pi}^+(p^2)=\frac{f_{D\pi}^+(0)}{(1-p^2/m^2_{D^*})
(1-\alpha_{D\pi}p^2/m^2_{D^*})}~, 
\end{equation}
where in addition to the $D^*$-pole a second pole is present at the
effective mass $m_{D^*}/\sqrt{\alpha_{D\pi}}$. The normalization 
$f_{D\pi}^+(0)$ is given by (\ref{fD0}) and the deviation in shape
from the single-pole form is described by the parameter
\begin{eqnarray}\label{paras1}
\alpha_{D\pi} = 1- \frac{2m_{D^*}f_{D\pi}^+(0)}
{f_{D^*}g_{D^*D\pi}} = 0.01^{+0.11}_{-0.07}~.
\end{eqnarray}
The LCSR (\ref{fplus}) and (\ref{srcoupl}) thus
predict $\alpha_{D\pi}$ to be consistent with zero,
that is complete dominance
of the lowest lying $D^*$ pole and negligible influence of 
higher poles and continuum states.
In other words, here the use of the two-pole parametrization (\ref{paramD})
does not introduce any significant model-dependence.
For the nominal values of the parameters, (\ref{paramD}) is plotted 
in Fig. \ref{fig7}. 
Comparing this figure with
Fig. \ref{fig1} one clearly sees
that the $D^*$-dominance in $f^+_{D\pi}$ 
is more pronounced than the $B^*$-dominance in $f^+_{B\pi}$.
This finding can be understood by considering the splitting
between the ground and excited vector meson states in the heavy quark
limit, e.g.,
\be
\frac{m_{D^{*'}}^2 - m_{D^*}^2}{m_{D^*}^2} = \frac{2(\Delta'-\Delta)}{m_c}
+ O(m_c^{-2}) ~,  
\label{excited}
\ee
where we have used the heavy quark mass expansion 
$m_{D^{*}}^2 = m_c^2+2m_c\Delta$, and similarly for 
$m_{D^{*'}}$. Since the hadronic scales $\Delta$ and $\Delta'$ are 
independent of the heavy quark mass, the relative splitting 
of the states in the $D^*$ system is expected to be larger than the
splitting in the $B^*$ system. Consequently, the heavier $B^*$ states 
are expected to 
contribute more to the dispersion relation (\ref{vectpole}) for 
$f_{B\pi}^+(p^2)$ 
than the excited $D^*$ states to the 
corresponding dispersion relation for 
$f_{D\pi}^+(p^2)$.  

\begin{figure}[p]
\centerline{
\epsfig{file=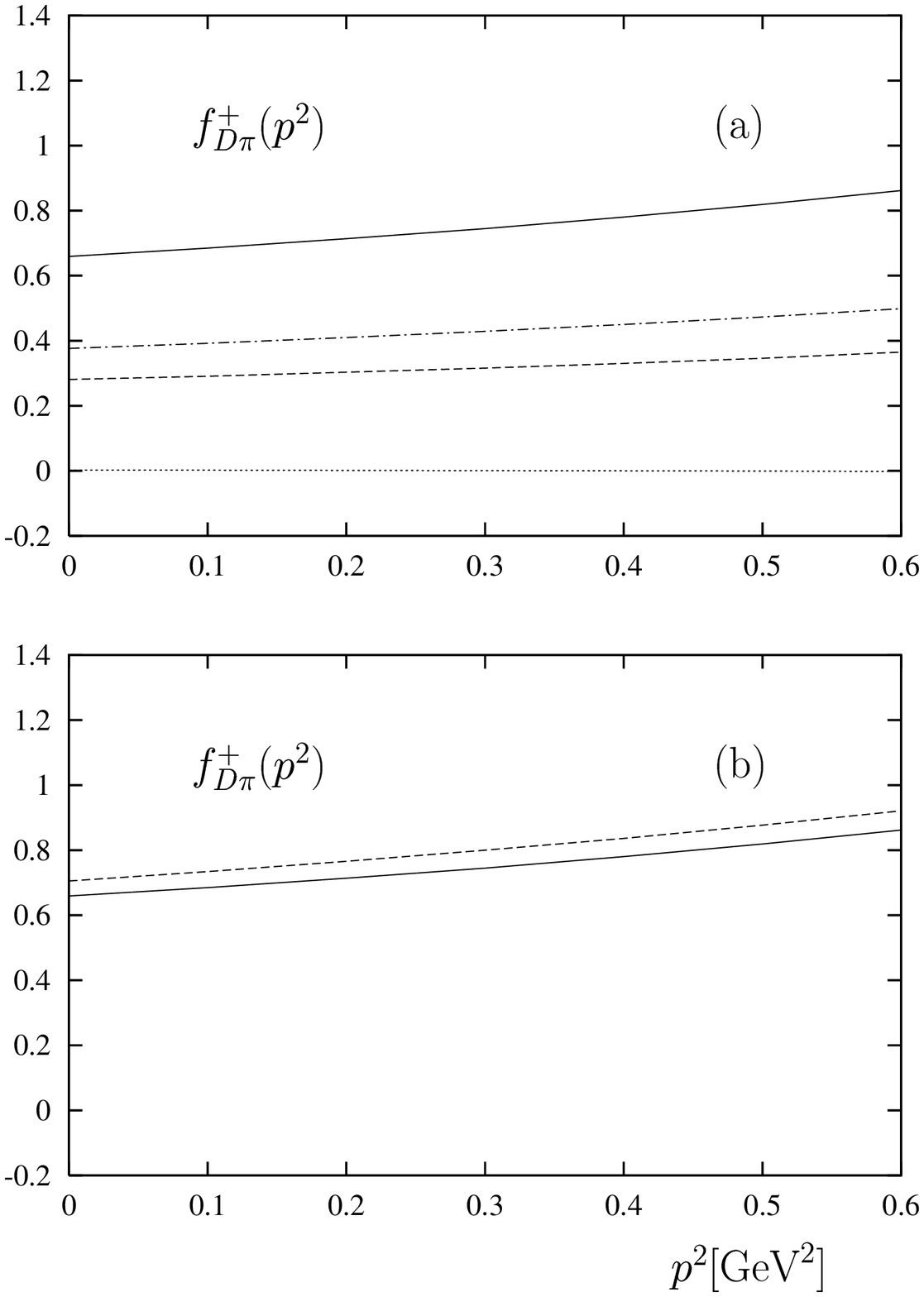,scale =0.6,width=\textwidth,bbllx=0pt,bblly=80pt,
bburx=600pt,%
bbury=800pt,clip=}
}
\caption{\label{fig8} \it The LCSR prediction on $f_{D\pi}^+$: 
(a) individual contributions
from twist 2 (dashed), 3 (dash-dotted), 4 (dotted), 
and the total sum (solid); 
(b) with (solid) and without (dashed) 
non-asymptotic corrections to the pion distribution amplitudes.} 
\end{figure}

The theoretical uncertainties in (\ref{fD0}) and (\ref{combinD})
are estimated analogously to the uncertainties in the $B$-meson case
explained in Sect. 3.
However, there are certain differences which should be pointed out.
Firstly, unlike in $B \rightarrow \pi$, the twist 3 term yields the 
largest individual contribution to $f_{D\pi}^+(p^2)$ as shown
in Fig. \ref{fig8}a. 
This causes no problem for the light-cone expansion itself 
because even and odd twists 
are associated with different chiral structures of the underlying 
correlation function. The corresponding series are, therefore,  
independent from each other. 
More definitely, the expansion of the term
in the heavy quark propagator proportional to the quark mass
gives rise to a series in even twist, whereas the term proportional 
to the quark momentum is expanded in a series of terms carrying 
odd twist.
Important for the convergence of the expansion is the observation that the 
twist 4 contribution is heavily suppressed with respect to the  
twist 2 contribution as shown in Fig. \ref{fig8}a. 
Unfortunately, very little is known about the twist 5 term. 
If twist 5 is similarly suppressed with respect to twist 3 
as twist 4 is suppressed with respect to twist 2, 
the uncertainty due to the neglected  
higher-twist contributions is practically negligible.
This assumption is made tacitly in all LCSR calculations. 
Secondly, the uncertainties in the nonasymptotic 
coefficients of the pion distribution amplitudes induce an almost 
momentum-independent uncertainty in the form factor $f^+_{D\pi}$ 
by about 5\% in contrast to the shape-dependent uncertainty
in $f^+_{B\pi}$. This can be clearly seen by comparing Fig. \ref{fig8}b 
and Fig. \ref{fig5}b.
Thirdly, for $D \rightarrow \pi$ 
the NLO QCD corrections to the LCSR at twist 2 
tend to be smaller than the corrections to the LCSR for $B \rightarrow \pi$,
in contradiction to naive expectation.  
The reason is that the shrinkage of $\alpha_s(m_Q)$ 
when going from charm to bottom 
is over-compensated by the growth of logarithms of the heavy quark mass
appearing in the coefficient functions of the sum rules \cite{KRWY2}.  
Quantitatively, the NLO effects on $f_D$, $f^+_{D\pi}$ and 
$f_{D^{*}}g_{D^*D\pi}$ amount to about 10\%, in contrast to 
$(20 \div 30)\%$ corrections in the $B$-meson case. We also expect this
tendency to persist for the missing NLO corrections at twist 3.
The maximal 15\% estimated for $B \rightarrow \pi$ would 
then correspond to about 6\% for $D \rightarrow \pi$. 

The $D\to K$ transition form factor is calculated from the LCSR
(\ref{fplus}) and (\ref{srcoupl}) using the same input parameters as for 
$D\to \pi$ with the exception 
of the distribution amplitudes. 
For the kaon distribution amplitude of twist 2 we take 
\begin{equation}
\varphi_K(u,\mu) = 6 u(1-u)\Big[1+ \sum_{n=1}^4a_n^{K}(\mu)C^{3/2}_n(2u-1)]
\label{expansion1}
\end{equation}
with $a_1^{K}(\mu_c)= 0.17$, $a_2^{K}(\mu_c)= 0.21$, $a_3^{K}(\mu_c)= 0.07$,
and $a_4^{K}(\mu_c)= 0.08$. In addition to the features 
of the pion distribution amplitude $\varphi_{\pi}(u,\mu)$ 
given in (\ref{expansion}), the above amplitude
incorporates effects from $SU(3)$-flavour violation. 
This is seen from the presence of the coefficients $a_{1,3}$ 
giving rise to asymmetric momentum distributions for the strange 
and nonstrange quark constituents inside the kaon.
The distribution amplitude  
$\varphi_K$ used in Ref. \cite{Ball} to calculate the $B\to K$ 
transition form factor differs from 
(\ref{expansion1}) in neglecting $a^K_{3,4}$ which, however, 
is quantitatively not important. 
Other $SU(3)$-breaking effects in the distribution amplitudes are neglected. 
This is certainly justified for the nonasymptotic terms of twist 3 and 4 
analyzed recently in Ref. \cite{BallDA}
which have anyway only a minor influence on the LCSR as pointed out earlier.
For a similar reason, the twist 3 and 4 quark-gluon 
distribution amplitudes are also taken in the $SU(3)$ limit. 
Furthermore, in the coefficients of the twist 2 and 3 distribution 
amplitudes $f_\pi$ is replaced by $f_K= 160$ MeV, and $\mu_{\pi}$
by $\mu_K=m_K^2/(m_s+m_{u,d})$, respectively. 
The latter brings the mass of the strange quark into the game
which is not very well known.
In Ref. \cite{BallDA} it was advocated to rely
on chiral perturbation theory in the $SU(3)$ limit and use
\be
\mu_K= \mu_\pi=
-2\frac{\langle \bar{q}q \rangle}{f_\pi^2} \,. 
\label{cpt}
\ee 
With the  quark condensate 
$\langle \bar{q}q \rangle(1 \mbox{GeV})=- (240 \pm 10~\mbox{MeV})^3$,
this yields   $m_s(1 \mbox{GeV})=150\pm 20$ MeV.  In the following, we take
$m_s(\mu_c)=150\pm 50$ MeV which covers the range of most
estimates including the rather low mass suggested by 
some of the lattice calculations \cite{revms}.

\begin{figure}
\centerline{
\epsfig{file=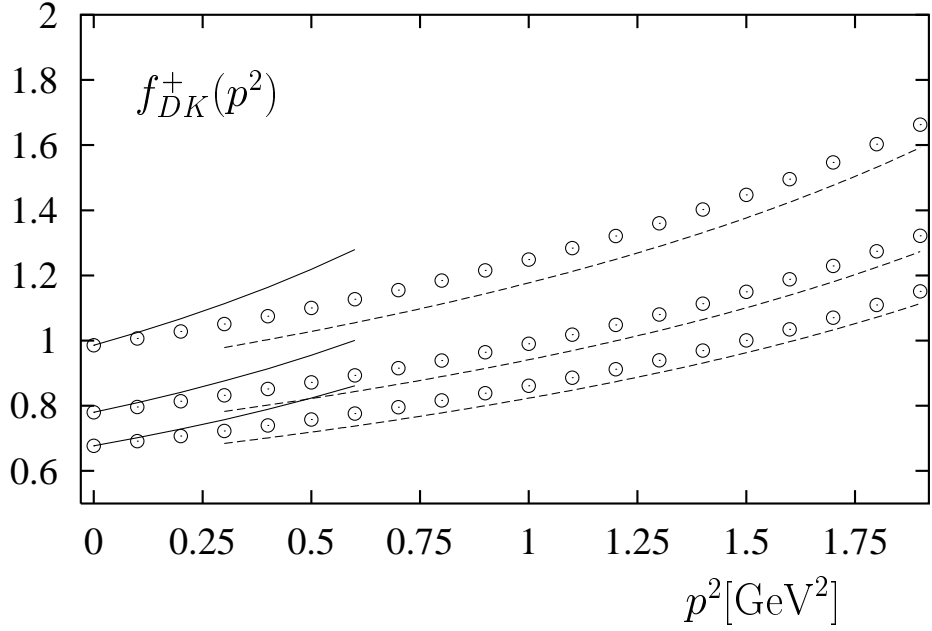,scale=0.8,%
clip=}}
\caption{\label{fig9} \it  The $D \to K$ form factor: direct 
LCSR prediction (solid curve),  
$D^*_s$-pole contribution with the LCSR estimate of the residue
(dashed curve), and two-pole parametrization (\ref{paramDK}) (circles).
Results are shown for $m_s =$ 100 MeV (upper part), 150 MeV 
(middle part) and 200 MeV (lower part).}
\end{figure}

The numerical predictions derived from the LCSR (\ref{fplus}) and 
(\ref{srcoupl}) with the above input are tabulated in 
Tab. 1 for three different values of the strange quark mass.
Fig. \ref{fig9} shows the form factor $f^+_{DK}$ at $p^2 < 0.6$ GeV$^2$
and the pole-contribution from the $D^*_s$ ground state as a function of
the momentum transfer squared.     
The two-pole parametrization analogous to (\ref{paramD}),
\begin{equation}\label{paramDK}
f_{DK}^+(p^2)=\frac{f_{DK}^+(0)}{(1-p^2/m_{D_s^*}^2)(1-
\alpha_{DK}p^2/m_{D_s^*}^2)}~,
\end{equation}
is also displayed in Fig. \ref{fig9}. For these plots we have chosen
the nominal values of the input parameters.
Whereas the normalization of $f^+_{DK}$ is rather 
sensitive to $m_s$ \cite{BKR}, the shape parameter
$\alpha_{DK}$ is more or less $m_s$-independent.  
Moreover, similarly as in the case of $f^+_{D\pi}$, $\alpha_{DK}$
is consistent with zero implying a strong dominance of the
$D^*_s$ pole.

\begin{table}[htb]
\caption{$D\to K$ transition parameters as a function of the $s$-quark mass}
\begin{center}
\begin{tabular}{|c|c|c|c|}
\hline
&&&\\

$m_s(\mu_c)$ ( MeV)
& 100  & 150 & 200 \\
&&&\\
\hline
&&&\\
$f^+_{DK}(0)$& $0.99\pm 0.11$ & $0.78\pm 0.11$ & $0.68\pm 0.09$\\ 
&&&\\
\hline
&&&\\
$f_{D^{*}_s}g_{D^*_s D K}$ & $3.9\pm 0.8$ &$3.1\pm 0.6$& $2.7\pm 0.5$ \\
&&&\\
\hline
&&&\\
$\alpha_{DK}$ & $-0.08^{+0.15}_{-0.07} $ & $-0.07^{+0.15}_{-0.07}$ 
& $-0.06^{+0.14}_{-0.07}$\\ 
&&&\\
\hline
\end{tabular}
\\
\end{center}
\end{table}

This is confirmed by measurements of 
$D\to K l^+ \nu $ in the CLEO \cite{CLEODK} and  
E687 \cite{E687} experiments. 
The data including those from earlier experiments 
are summarized in Ref. \cite{BaBar}. 
For illustration, fitting the single-pole formula 
\be
f^+_{DK}(p^2) = \frac{f^+(0)}{1-p^2/M^2}
\label{fDKpole}
\ee
to the world-average of the data one obtains 
\be
f^+(0)= 0.76 \pm 0.03, ~~M=2.00 \pm 0.15 \,\mbox{\rm GeV} ~.
\label{fDKexp}
\ee
While for $m_s \approx 150$ MeV the expectation on $f^+(0)$ and the 
experimental result perfectly match, the mass of the pole is
measured to be slightly lighter 
than $m_{D^*_s} = 2.11$ GeV, but within error it is still consistent 
with expectation.

There are also lattice determinations of the $D \to K$ form factor
which confirm the $D^*$-pole dominance.
The world average of $f^+_{DK}(0)$ is given by 
\cite{lattrev} 
\be
f^+_{DK}(0)= 0.73 \pm 0.07~, 
\label{fDKlat}
\ee
while a more recent calculation by the APE-Collaboration \cite{APE} yields 
$ f^+_{DK}(0)= 0.71 \pm 0.03 ^ {+0.00} _{-0.07}$.
Comparison of these results with Tab. 1 implies a 
$s$-quark mass which is even somewhat heavier than
150 MeV. While such a value is in accordance with 
the PCAC expectation (\ref{cpt}), it remains to be seen if it can
be reconciled with the 
smallest mass values resulting from lattice investigations \cite{revms}.

\section{ Semileptonic decay distributions and widths} 

\begin{figure}
\centerline{
\epsfig{file=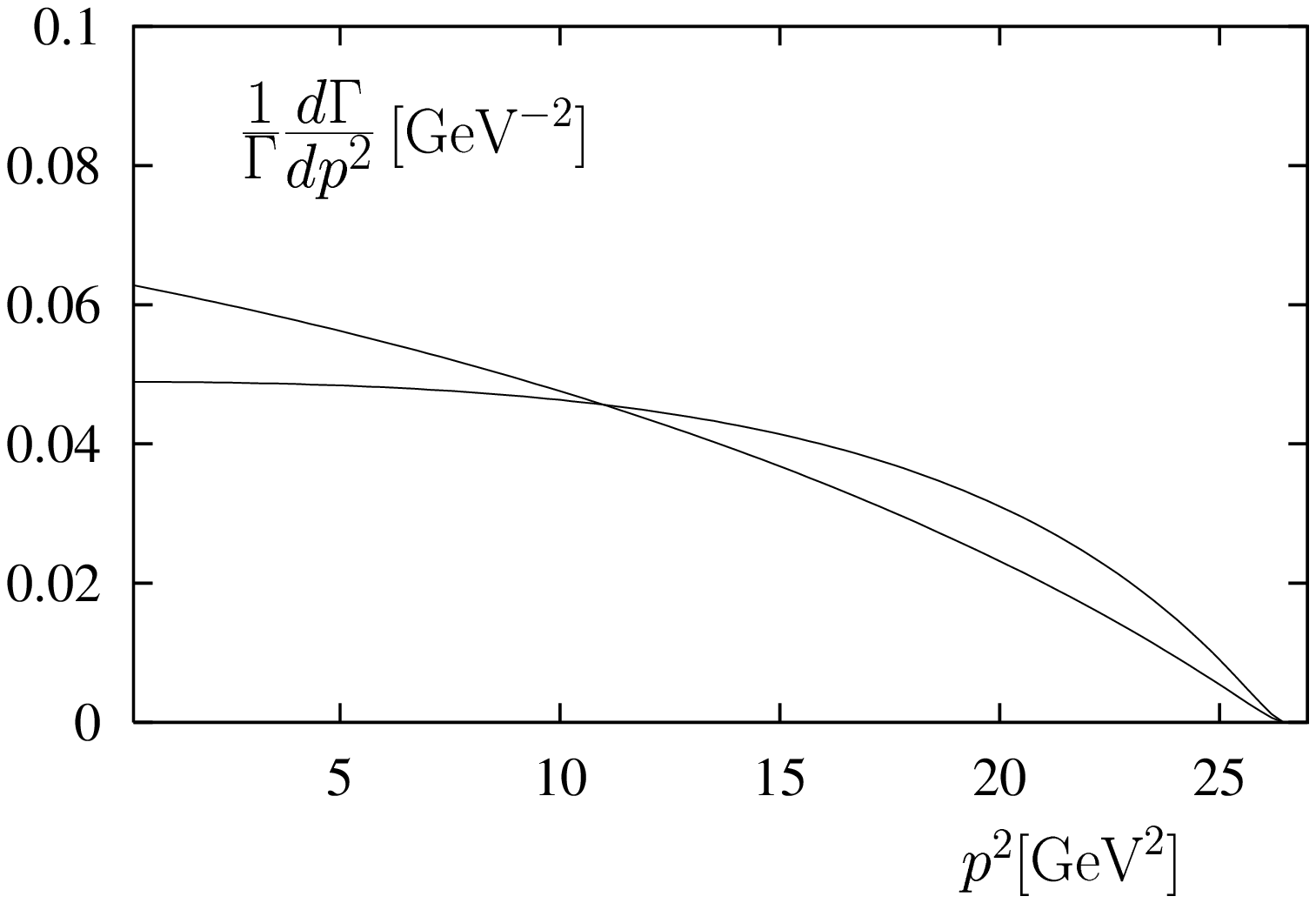,scale=0.8,%
clip=}}
\caption{\label{fig10} \it The normalized distribution of 
the momentum transfer squared in
$B \to \pi \bar l \nu_l$ ($ l=e, \mu$). 
The steeper distribution corresponds to $\alpha_{B\pi}=0.25$, 
the flatter to $\alpha_{B\pi}=0.53$.} 
\end{figure}

The parametrizations (\ref{paramB}), (\ref{paramD}) and (\ref{paramDK}) 
of the form factors $f^+_{B \pi}$, $f^+_{D \pi}$
and $f^+_{DK}$  can be used
to calculate the distributions of the momentum transfer squared $p^2$ 
in exclusive, semileptonic $B$ and $D$ decays, respectively. 
For $B\to \pi \bar{l}\nu_l$ with $l=e,\mu$ 
one has
\bq
\frac{d\Gamma( B\to \pi \bar{l} \nu_l)}{dp^2} = 
\frac{G^2|V_{ub}|^2}{24\pi^3}
(E_\pi^2-m_\pi^2)^{3/2}\left[f^+_{B\pi}(p^2)\right]^2 
\label{dG}
\eq
with $E_\pi= (m_B^2+m_\pi^2 -p^2)/2m_B$ being 
the pion energy in the $B$ rest frame.
The charged lepton is considered massless.
Substituting (\ref{paramB}) for $ f^+_{B\pi}(p^2)$ one obtains 
the integrated width
\be
\Gamma(B\to \pi \bar{l} \nu_l) 
= \int\limits_0^{(m_B-m_\pi)^2} dp^2 
\frac{d\Gamma(B\to \pi \bar{l} \nu_l)}{dp^2}
= (7.3 \pm 2.5) ~|V_{ub}|^2~\mbox{ps}^{-1}~.
\label{elmu}
\ee
It is important to note that
the theoretical uncertainty in the above mainly comes from the
uncertainty in the normalization parameter $f^+_{B\pi}(0)$. 
The uncertainty in the shape parameter $\alpha_{B\pi}$ 
has very little influence on the integrated width,
so does the use of the two-pole parametrization (\ref{paramB}).
This is anticipated from the normalized decay 
distribution plotted in Fig. \ref{fig10}.
Our result agrees with the integrated semileptonic width
\be
\Gamma(B\to \pi \bar{l} \nu_l) = 8.5 ^{+3.4}_{-1.5}
 ~|V_{ub}|^2~\mbox{ps}^{-1}
\label{elmulatt}
\ee
derived in Ref. \cite{Lellouch} from a  
lattice-constrained parametrization of $f_{B\pi}^+$. 

Experimentally, combining the branching ratio 
$BR(B^0\to\pi^-l^+\nu_l) = (1.8 \pm 0.6)\cdot 10^{-4}$
with the $B^0$ lifetime
$\tau_{B^0}=1.54 \pm 0.03$ ps \cite{PDG},   
one gets 
\be
\Gamma(B^0\rightarrow \pi ^- l^+ \nu_l) =
(1.17 \pm 0.39)\cdot10^{-4}~\mbox{ps}^{-1}~.
\label{CLEOpi7}
\ee
From that and (\ref{elmu})
one can then determine the quark mixing parameter $|V_{ub}|$.
The result is
\be
|V_{ub}|=(4.0 \pm 0.7 \pm 0.7)\cdot 10^{-3}
\label{result}
\ee
with the experimental error and theoretical uncertainty given in this order.
This value lies within the range of $|V_{ub}|$ quoted by the 
Particle Data Group \cite{PDG}:
\be
|V_{ub}|=0.002 \div 0.005~,
\label{resultPDG}
\ee  
and also agrees \cite{KR} with the determination 
from $B \to \rho \bar{l}\nu_l$ \cite{BB97}.

\begin{figure}
\centerline{
\epsfig{file=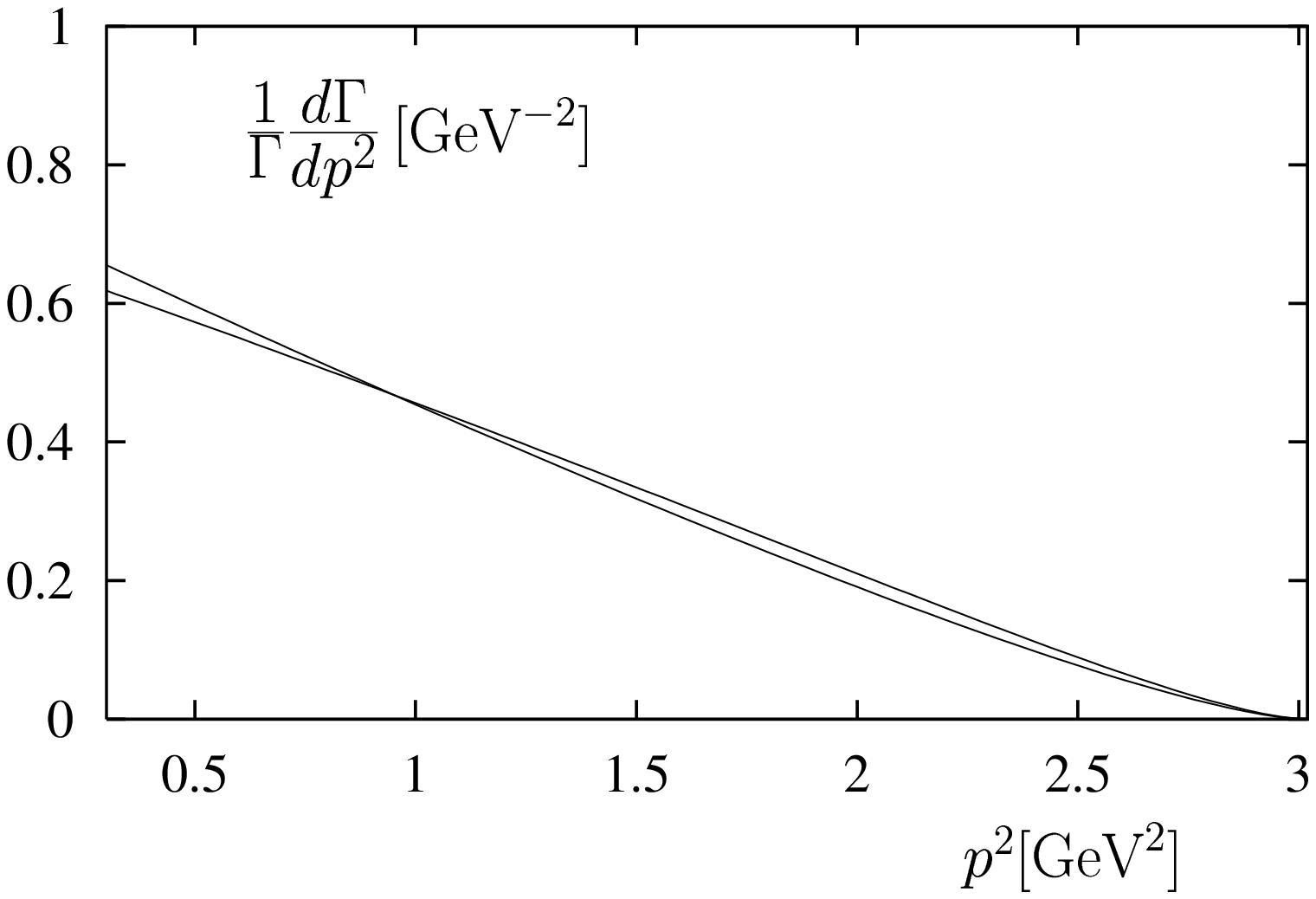,scale=0.8,%
clip=}}
\caption{\label{fig11} \it  The normalized distribution of 
the momentum transfer squared in
$D \to \pi \bar l \nu_l$ ($ l=e, \mu$). 
The steeper distribution corresponds to $\alpha_{D\pi}=-0.06$, 
the flatter to $\alpha_{D\pi}=0.12$.} 
\end{figure}

Conversely, turning to $D\to \pi \bar l \nu_l$ 
one can use the known value of $|V_{cd}|$
and subject the LCSR method to an experimental test.
From (\ref{paramD}) one obtains 
\be
\frac{\Gamma(D^0\rightarrow \pi ^- l^+ \nu_l)}{|V_{cd}|^2} = 
0.13 \pm 0.05 \quad \mbox{ps}^{-1}~,
\label{Dpilnu}
\ee
while the experimental width following from 
$BR(D^0\rightarrow\pi^- e^+ \nu_e) =
(3.7 \pm 0.6)\cdot 10^{-3}$,
$\tau_{D^0}= 0.415 \pm 0.004  ~\mbox{ps}$, 
and $|V_{cd}|= 0.224 \pm 0.016$ ~\cite{PDG}
is given by
\be
\frac{\Gamma(D^0\rightarrow\pi^- e^+ \nu_e)}{|V_{cd}|^2} =
0.177 \pm 0.038\quad \mbox{ps}^{-1}~.
\label{brD}
\ee
The agreement is satisfactory, although the 
theoretical and experimental uncertainties are still too big 
for a really decisive test. For this reason, it is particularly interesting 
to note the very small theoretical uncertainties in the
normalized momentum distribution shown in Fig. \ref{fig11}. 
It would be very useful to have comparably precise data.

\begin{figure}
\centerline{
\epsfig{file=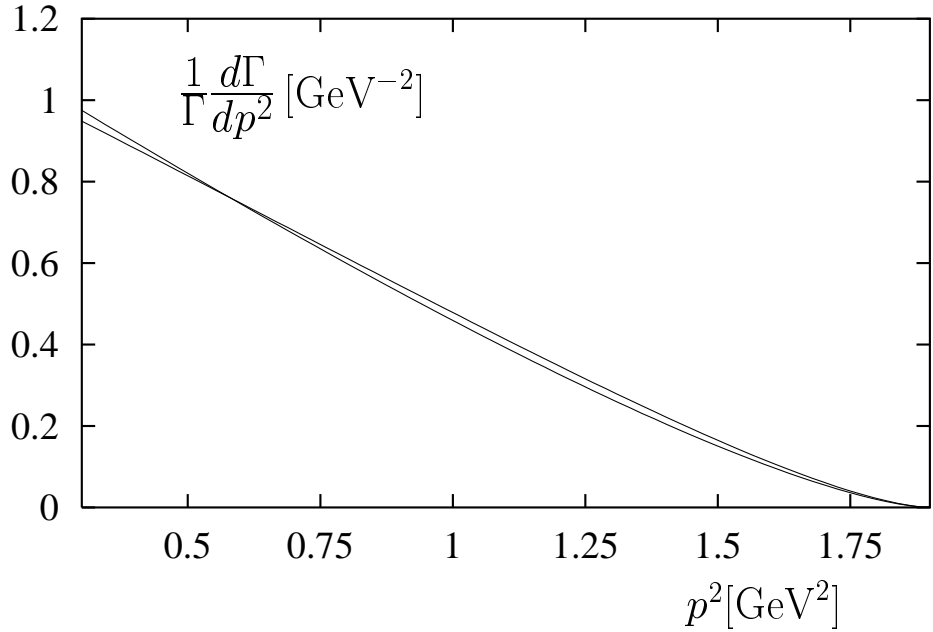,scale=0.8,%
clip=}}
\caption{\label{fig12} \it  The normalized distribution of 
the momentum transfer squared in
$D \to K \bar l \nu_l$ ($ l=e, \mu$). 
The steeper distribution corresponds to $\alpha_{DK}=-0.14$, 
the flatter to $\alpha_{DK}=0.08$.} 
\end{figure}

Similarly for $D \to K \bar l \nu_l$, 
(\ref{paramDK}) and Tab. 1 yield
\be
\frac{\Gamma(D^0\rightarrow K ^- l^+ \nu_l)}{|V_{cs}|^2}  = 
\left\{\begin{array}{c}
0.151 \pm 0.058 \quad \mbox{ps}^{-1},~~m_s(\mu_c)=100 ~\mbox{MeV},\\ 
0.094 \pm 0.036\quad \mbox{ps}^{-1},~~m_s(\mu_c)=150 ~\mbox{MeV},\\
0.072 \pm 0.027\quad \mbox{ps}^{-1},~~m_s(\mu_c)=200 ~\mbox{MeV},\\ 
\end{array}
\right.
\label{DKlnu}
\ee
which should be compared with the experimental width 
\be
\frac{\Gamma(D^0\rightarrow K^- l^+ \nu_l)}{|V_{cs}|^2} =
0.087 \pm 0.004 \quad \mbox{ps}^{-1}
\label{brDK}
\ee
derived from $BR(D^0\rightarrow K^- l^+ \nu) = (3.49 \pm 0.17)\%$ 
together with the above $D^0$ lifetime,
and $|V_{cs}|= 0.9734 \div 0.9749$ \cite{PDG}. The latter
interval for  $|V_{cs}|$ obtained from unitarity of the CKM matrix 
is very tight and has therefore a negligible influence on 
the experimental error in (\ref{brDK}).
Despite of the considerable uncertainty remaining even when $m_s$
is fixed, the comparison of LCSR and data clearly favours
a relatively heavy strange quark
mass of about 150 MeV or slightly more.
This conclusion is confirmed
by considering the ratio of partial widths 
\be
R=\frac{\Gamma(D^0\rightarrow \pi ^- l^+ \nu_l)}{ 
\Gamma(D^0\rightarrow K ^- l^+ \nu_l)}= 
\left\{\begin{array}{c}
0.04, ~~m_s=100 ~\mbox{MeV},\\ 
0.07,~~m_s=150 ~\mbox{MeV}, \\
0.10,~~m_s=200 ~\mbox{MeV}\\ 
\end{array}
\right. 
\label{ratio}
\ee
in which the uncertainties not related to the strange quark mass
drop out to a large extent.
The experimental world average quoted in \cite{BaBar},
\be
R = 0.10 \pm 0.02,
\label{rexp}
\ee
again supports a strange quark mass between 150 MeV and 200 MeV. 
Contrary to the integrated width,
the distribution in momentum transfer is insensitive to $m_s$.
Moreover, since the shape parameter $\alpha_{DK}$ is very small or
vanishing, the normalized decay distribution in 
$D \to K \bar l \nu_l$ is predicted 
very reliably as demonstrated in Fig. \ref{fig12}. It can therefore serve as a 
stringent test similarly as the corresponding distribution in
$D \to \pi \bar l \nu_l$.

We conclude with a final remark on the uncertainties in the LCSR results.
Unlike phenomenological quark models, the LCSR approach allows 
to estimate the uncertainty in a given result within the same framework. 
This is one of the main virtues of QCD sum rules. 
The model-dependence due to the use of a particular
two-pole parametrization of the form factors is negligible or
unimportant in the present applications. In $B \to pi$ it only 
concerns the region of large momentum transfer, and is therefore
negligible in the integrated semileptonic width and the value
of $|V_{ub}|$ extracted from it. The $D$ meson form factors,
on the other hand,  
are completely dominated by the pole of the respective ground-state
vector meson, and therefore insensitive to the effective second pole
modelling effects from excited states.
In total, the present theoretical uncertainty 
in the integrated widths of the
semileptonic $B$ and $D$ decays considered in this paper
is estimated to be about 30 to 40 \%.

Thus, in order to match the accuracy of the data on  
$D \to \pi \bar l \nu_l$ and the precision of 
$B \to \pi \bar l \nu_l$ measurements expected 
at the new $B$ factories,
the theoretical 
uncertainties have to be reduced by more than a factor of 2. This
is not impossible, but will require considerable efforts. 
Among the most important tasks are the NLO calculation of the twist 3 
contributions, a reanalysis of the non-asymptotic corrections to the
light-cone distribution amplitudes including mass effects, 
and an at least rough estimate of the size of the twist 5 terms.    
 
\bigskip 

{\bf Note added}

For completeness, we also present an update  
of the LCSR prediction on the $B\to K$ transition form factor 
$f^+_{BK}$. This form factor  
plays an important role in the theoretical analysis 
of the rare decays $B\to K l^+l^-$, $l=e,\mu,\tau$ \cite{AliBall}. 
The previous  LCSR estimates of $f^+_{BK}$ \cite{BKR,Ball} 
are restricted to the region of small and intermediate 
momentum transfer and not applicable to large 
momentum transfer where the $B_s^*$ pole contribution 
with the residue given by the $B_s^*B K$ coupling
becomes dominant. Here, we have performed an analysis in analogy 
to the calculation of the $B\to \pi$ form factor described in 
this paper with the pion distribution amplitudes replaced by the
corresponding 
kaon distribution amplitudes given in section 4. 
Calculating $f^+_{BK}(0)$ from the LCSR (\ref{fplus})
and $f_{B^*_s}g_{B_s^*BK}$ from the LCSR (\ref{srcoupl}),
and deriving, from that, the remaining parameter $\alpha_{BK}$ 
of the two-pole parametrization 
\be\label{paramBK}
f_{BK}^+(p^2)=\frac{f_{BK}^+(0)}{(1-p^2/m_{B^*_s}^2)(1-
\alpha_{BK}p^2/m_{B^*_s}^2)} ~, 
\ee
we find $f^+_{BK}(0) =0.36 \pm 0.07 (0.33\pm 0.06; 0.41 \pm 0.08)$ 
for $m_s=150 (200;100) $ MeV and  $\alpha_{BK}= 0.28^{+0.29}_{-0.08}$.
The ground-state mass  $m_{B^*_s}= 5.416$ GeV 
is taken from \cite{PDG}. 

\bigskip 

{\bf Acknowledgements}

We are grateful to P. Ball and V. M. Braun for useful discussions. 
This work was supported by the Bundesministerium f\"ur Bildung und
Forschung (BMBF) under contract number 05 HT9WWA 9.
In addition, O.Y. acknowledges support from the 
US Department of Energy (DOE), and A.K. wishes to thank the 
Danish Research Council for Natural Sciences and the 
Niels Bohr Institute for support.
S.W. is grateful for the hospitality and support
during his visit at the University of W\"urzburg.

\newpage

\end{document}